\documentclass[%
superscriptaddress,
showpacs,preprintnumbers,
amsmath,amssymb,
 aps,
prb,
twocolumn,
10pt,
]{revtex4-1}

\usepackage{graphicx}
\usepackage{dcolumn}
\usepackage{bm}
\usepackage{hyperref}
\usepackage{textcomp}
\usepackage{siunitx}
\mathchardef\mhyphen="2D

\begin{document}


\title{Comparison between thermal and current driven spin-transfer torque in nanopillar metallic spin valves}


\author{J. Flipse}
\email[]{J.Flipse@rug.nl}
\author{F. K. Dejene}
\author{B. J. van Wees}
\affiliation{Physics of Nanodevices, Zernike Institute for Advanced Materials, University of Groningen, Nijenborgh 4,
9747 AG Groningen, The Netherlands.}


\date{\today}


\begin{abstract}
We investigate the relation between thermal spin-transfer torque (TSTT) and the spin-dependent Seebeck effect (SDSE), which produces a spin current when a temperature gradient is applied across a metallic ferromagnet, in nanopillar metallic spin valves. Comparing its angular dependence (aSDSE) with the angle dependent magnetoresistance (aMR) measurements on the same device, we are able to verify that a small spin heat accumulation builds up in our devices. From the SDSE measurement and the observed current driven STT switching current of 0.8 mA in our spin valve devices, it was estimated that a temperature difference of 230~K is needed to produce an equal amount of TSTT. Experiments specifically focused on investigating TSTT show a response that is dominated by overall heating of the magnetic layer. Comparing it to the current driven STT experiments we estimate that only $\sim$10\% of the response is due to TSTT. This leads us to conclude that switching dominated by TSTT requires a direct coupling to a perfect heat sink to minimize the effect of overall heating. Nevertheless the combined effect of heating, STT and TSTT could prove useful for inducing magnetization switching when further investigated and optimized.
\end{abstract}

\maketitle
\section{Introduction} \label{Intro}
In spintronics the intrinsic angular momentum of the electron (spin) is used to develop new or improved electronic components. In the spin-transfer torque (STT) mechanism proposed by Slonczewski and Berger in 1996 \cite{Slon_STT, Berger_STT}, a spin polarized charge current entering a magnetic layer exerts a torque on the magnetization by transfer of angular momentum. Nowadays STT is being extensively studied and STT switchable random access memory (STT-RAM) is one of the prime candidates for replacing dynamic RAM (DRAM) in the future \cite{STT_switch}. The two spin channel model \cite{two_channel} describes collinear transport, in for instance giant magnetoresistance devices, but is not able to explain and quantify the absorption of transverse spins in STT. Therefore a so called spin mixing conductance ($G_{\uparrow \downarrow}$) was defined \cite{MixC2, MixC} that gives the efficiency with which these spins transverse to the magnetization direction are absorbed at the non-magnetic (N)$|$ferromagnetic (F) interface. $G_{\uparrow \downarrow}$ can be determined experimentally by performing angular magnetoresistance measurements.\cite{aMR, aMR_exp}

In recent years research in the field of spin caloritronics, the interplay between spin and heat transport, has led to exciting new results.\cite{spin_calor} In the spin-dependent Seebeck effect (SDSE) \cite{SDSE_1, SDSE_2} heat flow is used to inject a spin polarized current from F into N, which can exert an STT on the magnetization of a second F layer. Indications of such a TSTT have been reported by Yu et al. \cite{TT}, where they observed a change in the switching field of a Co$|$Cu$|$Co spin valve in the second harmonic response to a current sent through the nanowire. Nevertheless a complete study where the efficiency of the TSTT is quantified and a comparison with STT is made, is still lacking.

The goal of this paper is to provide such a study of TSTT in F$|$N$|$F GMR nanopillars. Using the same device to study the GMR, the SDSE as well as their angle dependence we are able to reliably compare both. Furthermore we discuss measurements oriented at directly observing TSTT and the obstacles that come with it.   

This paper is organized as follows. In section \ref{sec:aMR_aSSE}, we discuss the theory of STT and TSTT and specifically describe how the angle dependent GMR and SDSE in magneto electron circuit theory provides a way to quantify both mechanisms. Furthermore we show that a spin heat accumulation affects the aSDSE measurements, as the energy dependence of the spin mixing conductance becomes relevant. Section \ref{sec:exp} describes the device fabrication as well as the measurement techniques that were used. Section \ref{sec:exp_results} presents the GMR and SDSE measurement results and compares their angle dependences. The difference between the two leads us to conclude that a spin heat accumulation builds up in our devices. Section \ref{sec:TT} presents measurements where the effect of TSTT on the magnetic switching field is studied. In section \ref{sec:disc} we discuss the results presented and conclude that only $\sim$10\% of the response is due to TSTT.

\begin{figure*}[t]
\includegraphics{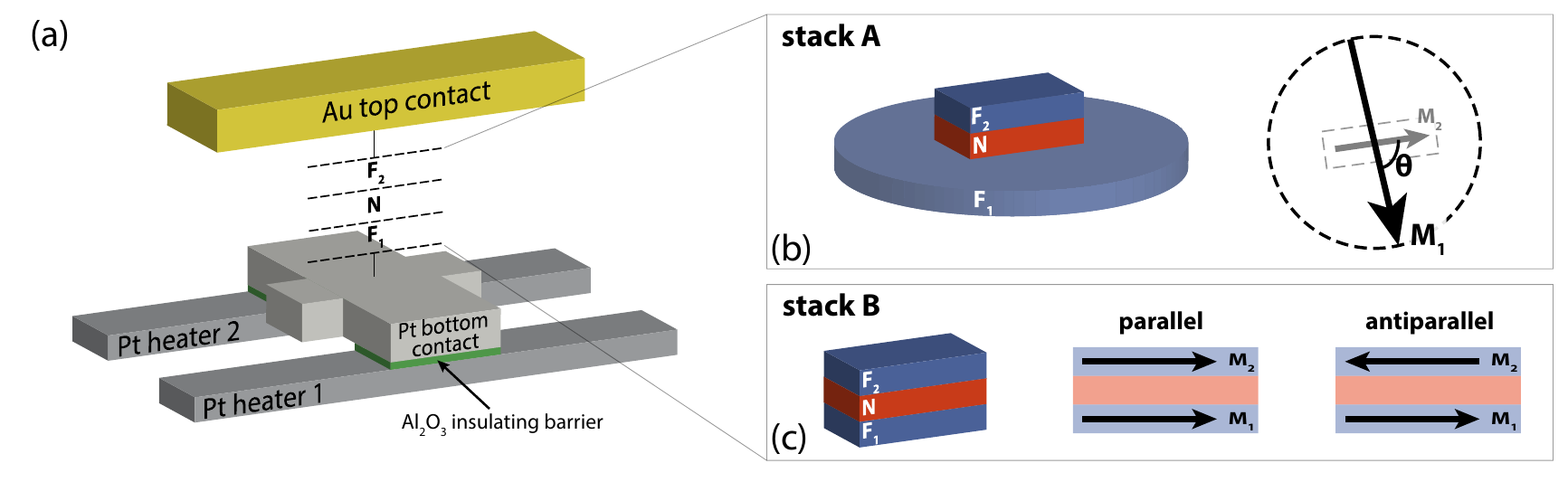}
\caption{(Color online) (a) Schematic representation of the device structure used, where an F$|$N$|$F stack is sandwiched between a Pt bottom contact and a Au top contact. Two Pt Joule heaters are used to produce a thermal gradient across the stack and are insulated form the Pt bottom contact by a thin Al$_{2}$O$_{3}$ layer. (b) In the angle dependent measurements stack type A is used, consisting of a circular F$_{1}$ layer and rectangular N and F$_{2}$ layers. The circular shape of F$_{1}$ ensures that there is no preferential in plane direction for the magnetization, such that it easily aligns with a small magnetic field. Rotating this small field will not influence the magnetization direction of F$_{2}$ giving an angle $\theta$ between M$_{1}$ and M$_{2}$. (c) For the thermal STT measurements stack type B is used, consisting of in situ grown rectangular F$|$N$|$F stack. Because of shape anisotropy two stable magnetic states are present, namely parallel and anti parallel magnetization alignment.}
\label{fig:sample}
\end{figure*}

\section{Theory} \label{sec:aMR_aSSE}
If, in an F$|$N$|$F stack, the magnetization of one of the F layers is rotated while keeping the other pinned, noncollinear spin transport becomes important. The spin current flowing from one F layer to the other will have a spin component transverse to the magnetization direction of the second F layer. Contrary to the collinear case these transverse spin components are not eigenstates of the ferromagnet and its angular momentum will be absorbed by destructive interference in F over the decoherence length, expected to be $\leq$1 nm for transition metals.\cite{des_intf} The absorbed angular momentum gives a torque on the magnetization which, if large enough, can excite magnetization dynamics or even reverse its direction. In magnetoelectronic circuit theory \cite{MixC} the real part of the spin mixing conductance ($G^{r}_{\uparrow \downarrow}$), in typical metals an order of magnitude larger than the imaginary part, gives the efficiency with which the electron's spin component transverse to the magnetization (M) direction are absorbed at an F$|$N interface:\cite{MixC}
\begin{equation}
I_{s,\perp} = V_{s,\perp} G^{r}_{\uparrow \downarrow}
\label{eq:spin_mixing}
\end{equation}
where $I_{s,\perp}$ is the transverse angular momentum current absorbed and $V_{s,\perp}$ is the the spin accumulation ($V_{\uparrow}-V_{\downarrow}$) at the F$|$N interface with the electron spin pointing perpendicular to M. The charge current through an F$|$N$|$F stack \cite{des_intf} depends on the angle between the two magnetizations ($\theta$) in the thermalized regime as follows:\cite{aSDSE}
\begin{widetext}
\begin{equation}
I_{c} (\theta) = \frac{G}{2} \left[ \left(1 - \frac{P_{G}^{2}\tan^{2}(\theta/2)}{\eta + \tan^{2}(\theta/2)} \right) \Delta V + \left( 1- \frac{P_{G}P'\tan^{2}(\theta/2)}{\eta + \tan^{2}(\theta/2)}\right) S\Delta T \right]
\label{eq:current}
\end{equation}
\end{widetext} 
where $\Delta V$ and $\Delta T$ are the voltage and temperature difference across the spin active part of F \cite{active_part}, $S$ is the F's Seebeck coefficient or thermopower, $P_{G}$=$\left(G_{\uparrow} - G_{\downarrow}\right) / G$ is the spin polarization of the F's conductance, $P'$=$\left(P_{S}+2P_{G}-P_{S}P_{G}^{2}\right) / 2$ with the spin polarization of the Seebeck coefficient $P_{S}$=$\left(S_{\uparrow} - S_{\downarrow}\right)/{S}$ and $\eta=2G^{r}_{\uparrow \downarrow}/ G$ with $G=G_{\uparrow}+G_{\downarrow}$ \cite{aSDSE}.

The $G^{r}_{\uparrow \downarrow}$ can be determined for a certain F$|$N interface by using $\eta$ as a fitting parameter for angle dependent magnetoresistance (aMR) measurements, by setting $\Delta T=0$ in Eq. \ref{eq:current} (see appendix \ref{app_derivation}):\cite{aMR, aMR_exp}
\begin{equation}
\text{aMR} = \frac{R(\theta)}{R(0)} = \frac{\eta + \tan^{2}(\theta/2)}{\eta + (1-P_{G}^{2})\tan^{2}(\theta/2)}
\label{eq:aMR}
\end{equation} 

A similar approach can be used for the angle dependence of the SDSE (aSDSE), which is given by Eq. \ref{eq:current} for $I_{c}=0$ (see appendix \ref{app_derivation}):\cite{aSDSE}   
\begin{equation}
\text{aSDSE} = \frac{-\Delta V(\theta)}{S\Delta T} = \frac{\eta + \left(1-P_{G}P'\right)\tan^{2} \left(\theta/2\right)}{\eta + \left(1-P_{G}^{2}\right)\tan^{2} \left(\theta/2\right)}
\label{eq:aSdSE}
\end{equation}

Both the MR and the SDSE produce a spin current running from one of the F layers to the other and therefore lead to a spin transfer torque, either current driven (STT) or driven by a temperature difference (TSTT).
\begin{align}
\tau_{STT} (\theta) &= \frac{\hbar}{2e}A(\theta) P_{G}I_{c} \label{eq:STT}\\[12pt]
\tau_{TSTT} (\theta) &= \frac{\hbar}{2e}\frac{G}{2} A(\theta) \left(P'-P_{G}\right)S\Delta T \label{eq:TSTT}
\end{align}
with \vspace{0.5em} $A(\theta)=\frac{\eta \sin(\theta)}{\eta \left(1+\cos (\theta) \right) + \left(1-\cos (\theta) \right)} \frac{\eta+\tan^{2}\left(\theta/2\right)}{\eta+\left(1-P_{G}^{2}\right)\tan^{2}\left(\theta/2\right)}$. 

The description given above holds in the thermalized regime where strong inelastic scattering between the two spin channels leads to energy exchange and ensures that they remain at the same temperature. However if inelastic scattering is relatively weak the electron temperatures can become spin-dependent and a spin heat accumulation will build up \cite{spin_heat, spin_heat_exp}. Such a spin heat accumulation produces an additional SDSE term which depends on the spin heat accumulation itself and the energy derivative of $G^{r}_{\uparrow \downarrow}$, and a normalized spin mixing thermopower can be defined $\eta'=2\left(\delta G^{r}_{\uparrow \downarrow}/\delta E\right)_{E=E_{F}}/ \left(\delta G/\delta E\right)_{E=E_{F}}$ \cite{aSDSE}. As a consequence the aSDSE curve shape will differ from that in the thermalized regime and Eq. \ref{eq:aSdSE} will not accurately describe the observed aSDSE behaviour.

\begin{figure*}[t]
\includegraphics{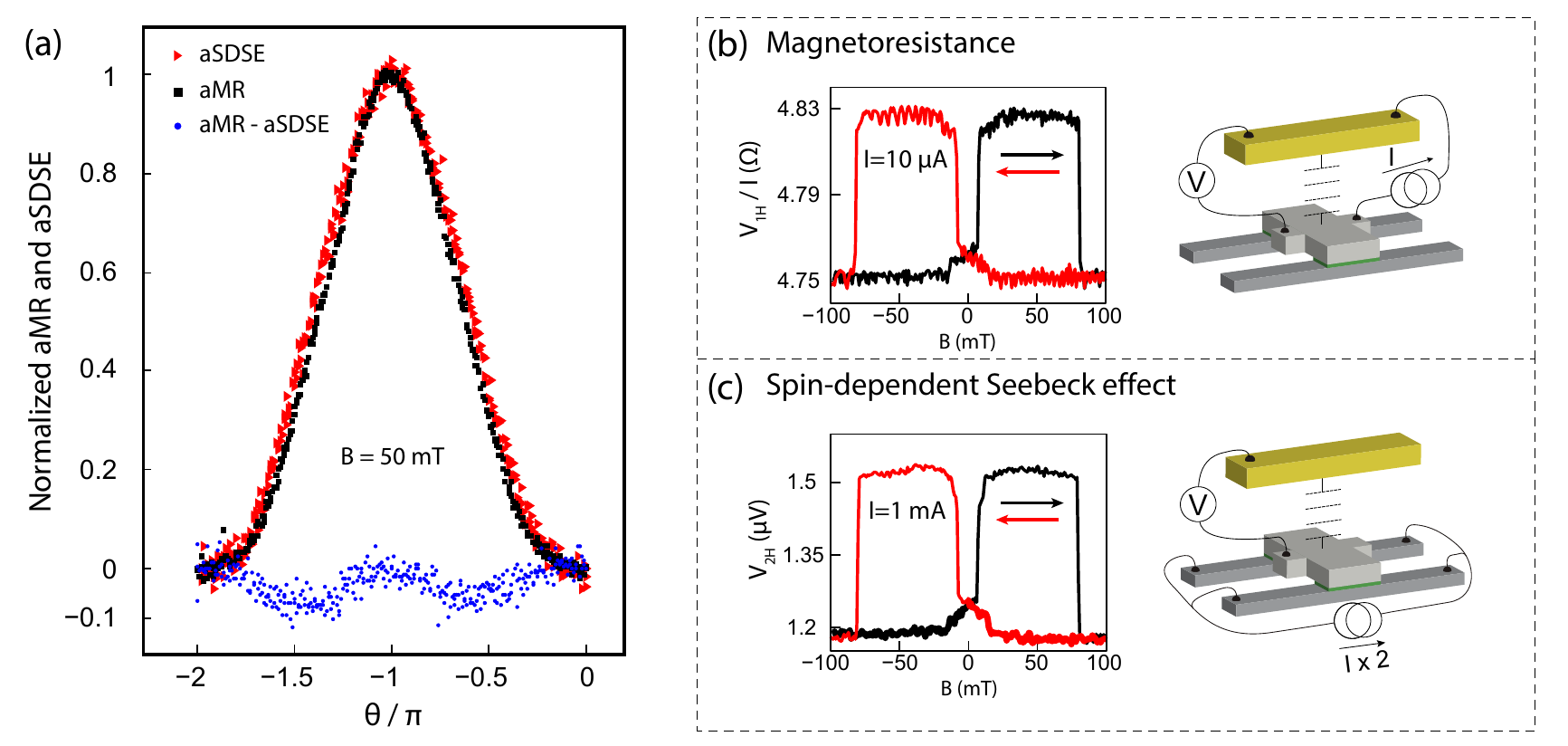}
\caption{(Color online) (a) The normalized, angle dependent magnetoresistance (aMR) \cite{aMR_norm} (squares), angle dependent spin-dependent Seebeck effect (aSDSE) (triangles) and the difference between the two (circles) are plotted as a function of the angle between M$_{1}$ and M$_{2}$. (b) The magnetoresistance (MR) measurement gives the resistance across the stack as a function of applied magnetic field B. (c) The spin-dependent Seebeck effect gives the Seebeck voltage as a function of applied magnetic field B.}
\label{fig:angle_meas}
\end{figure*}

\section{Fabrication and measurement techniques} \label{sec:exp}
The samples are prepared on top of a thermally oxidized Si substrate by 8 or 9 consecutive electron-beam lithography (EBL) steps, depending on the stack type. All the materials are deposited by e-beam evaporation with a base pressure of $3\times 10^{-6}$ mbar.

In this paper two types of F$|$N$|$F stacks are used. One for the angle dependent measurements (section \ref{sec:exp_results}) and an other for the TSTT measurements (section \ref{sec:TT}), for convenience they are named stack type A (Fig. \ref{fig:sample}(b)) and B (Fig. \ref{fig:sample}(c)) respectively. For both stack types the full device consists of a bottom platinum (Pt) contact of 60 nm thick and a 130 nm thick gold (Au) top contact with the F$|$N$|$F stack sandwiched in between (see figure \ref{fig:sample}(a)). On both sides of the Pt bottom contact Pt Joule heaters of 40 nm thick are placed to produce a thermal gradient across the F$|$N$|$F stack. An 8 nm thick aluminium oxide (Al$_2$O$_{3}$) layer seperates these Pt heaters from side extensions of the Pt bottom contact, ensuring strong thermal contact between the two but excluding any direct electrical pick up. Around the stack a $\sim$150 nm thick layer of polymethyl methacrylate (PMMA) e-beam resist is crosslinked, electrically isolating the top from the bottom contact.

The stack of type A, see figure \ref{fig:sample}(b), is fabricated in two steps. First a circular F$_{1}$ layer of 15 nm thick Permalloy (Py)(Ni$_{80}$Fe$_{20}$) is deposited, with a diameter of 300 nm. After cleaning the interface by Ar ion milling, to create a good Ohmic contact, the remainder of the stack is deposited consisting of a 10 nm thick copper layer (Cu) followed by a 10 nm thick Py layer with lateral dimensions of $150\times50$ nm$^{2}$. Because F$_{1}$ is circular there is no preferential in-plane direction for the magnetization. The magnetization of F$_{1}$ (M$_{1}$) will therefore easily follow a relatively small rotating applied magnetic field. However, the rectangular shape of F$_{2}$ ensures an easy axis for M$_{2}$, parallel to its longest side, due to shape anisotropy. Therefore the rotation of M$_{2}$ is negligible when the applied field is much lower than the field needed to rotate M$_{2}$ or to overcome its hard axis direction. Such magnetic behavior is ideal for magnetization angle dependent measurements, further discussed in section \ref{sec:exp_results}.

Stacks of type B, see figure \ref{fig:sample}(c), are rectangular in shape ($100\times50$ nm$^{2}$) and consists of 15 nm (F$_{1}$) and 5 nm (F$_{2}$) thick Py layers separated by a 15 nm thick Cu spacer. The full stack is deposited without breaking vacuum. Both magnetic layers have the same easy axis direction giving two distinct stable states, namely parallel or anti parallel alignment of the two magnetizations. These stacks are used in section \ref{sec:TT} to investigate changes in switching field due to TSTT.

The electrical measurements presented in this paper are all performed using standard lock-in detection techniques, providing a way to distinguish first harmonic response signals ($V_{1\text{H}}\propto I$) from second harmonic response signals ($V_{2\text{H}}\propto I^{2}$). To ensure a thermal steady-state condition a low excitation frequency of 17~Hz was used. All measurements are performed at room temperature except for the temperature dependent measurement in section \ref{sec:TT}, where a Peltier heating element together with a thermometer is used to bring and keep the sample at a preset elevated temperature.

\section{Angle dependent experiments} \label{sec:exp_results}
To investigate the aMR and aSDSE an F$|$N$|$F stack of type A is used (see Section \ref{sec:exp}). For characterization purposes we first measure the MR and SDSE.

Fig. \ref{fig:angle_meas}(b) gives the MR measurement where the resistance across the stack is measured as a function of the applied magnetic field (B), parallel to the easy axis of F$_{2}$. Just after B passes through zero the magnetization in the F$_{1}$ layer switches as it has no easy axis direction. The field necessary to switch the magnetization of F$_{2}$ is significantly larger, around 80 mT, as it has to overcome the planar shape anisotropy. Nevertheless the field to switch F$_{2}$ is larger than expected for a single layer of its size and shape. This is caused by the dipole magnetic field created by the F$_{1}$ layer coupling to the magnetization of F$_{2}$. For the angle dependent measurements we have to make sure that this coupling is canceled out such that it will not influence M$_{2}$ when rotating M$_{1}$. From separate measurements we conclude that the coupling corresponds to a 50 mT field, see appendix \ref{app_dipole}. A constant B of 50 mT in the angle dependent experiments is therefore sufficient to cancel out the dipole coupling field. Note that because we compare aMR and aSDSE measurements directly, measured on the same sample and using the same technique, any small differences between the angle set by the rotation of B and the actual angle, between M$_{1}$ and M$_{1}$, has no effect on the ability to compare both curves.

The MR measurement corresponds well with the results found from a Comsol Multiphysics three-dimensional finite element model (3D-FEM) with $P_{\sigma,1}=0.25$ and $P_{\sigma,2}=0.52$. See Refs. \onlinecite{spin_heat_exp, model} for a full discussion of the model. The difference in P$_{\sigma}$ for the two F layers is because of the ion mill cleaning of the F$_{1}$ layer,\cite{SDSE_1, SDSE_2} which leads to a stronger spin scattering and can thus be taken as an effective lower P$_{\sigma}$.  

The SDSE measurement in Fig. \ref{fig:angle_meas}(c) gives the Seebeck voltage measured across the stack while sweeping B. The temperature gradient over the stack is produced by sending a 1 mA root mean square current through each Pt Joule heater. A clear difference in the Seebeck voltage for the parallel and antiparallel case is observed. The SDSE signal and the background voltage correspond well with previously reported results \cite{SDSE_1, SDSE_2} and with the modeled values, with $P_{S,1}=0.19$ and $P_{S,2}=0.35$.

For the angle dependent measurements the sample holder is mounted on a rotatable stage with a rotation precision of at least $\pi/180$ radian by the automated control of a stepper motor. The sample holder is rotated from $-2\pi$ to $2\pi$ radian with a constant B of 50 mT while recording the voltage across the stack. M$_{1}$ will follow B therefore creating an angle $\theta$ between M$_{1}$ and M$_{2}$, see figure \ref{fig:sample}(b), equal to the rotation of the sample holder.

In Fig. \ref{fig:angle_meas}(a) the aMR and aSDSE measurements are plotted together and are normalized by the spin signal from the MR and SDSE measurements, respectively. In this way the angle dependence of both effects can directly be compared. A small but distinct difference between the two curves is visible, as the aSDSE is wider than the aMR, indicating that $\eta'$ starts playing a role. From this we can conclude that a SHA builds up in our stacks, verifying previous results of direct SHA measurements.\cite{spin_heat_exp} The TSTT, as described in Eq. \ref{eq:TSTT}, will be affected as well but from the relatively small difference between the aMR and aSDSE curves, of maximum 10\% of the total spin signal (see Fig. \ref{fig:angle_meas}(a)), we can assume that this change will be small and in first order can be neglected.

\section{Investigation of thermal spin-transfer torque} \label{sec:TT}

The existence of an SDSE suggests that the spin current generated by a thermal gradient across an F$|$N$|$F stack would produce a TSTT. The experiments discussed in this section are aimed at finding evidence for such a TSTT. For this purpose we use devices with F$|$N$|$F stack of type B (see Fig. \ref{fig:sample}(c)) to investigate the changes in minor loop switching fields. 

\begin{figure}[t]
\includegraphics{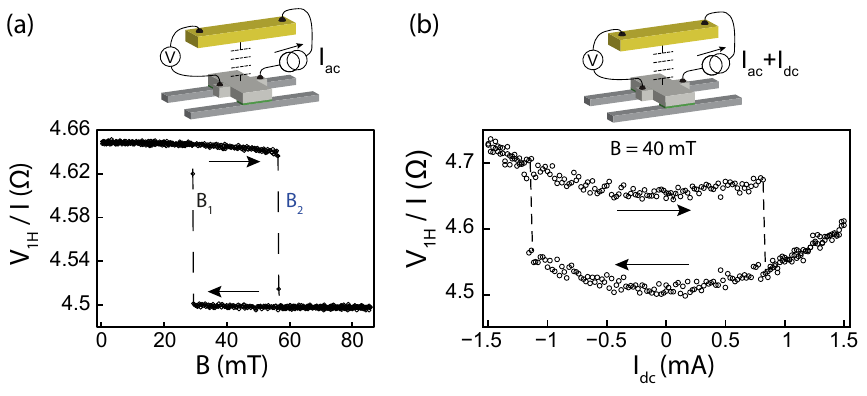}
\caption{(Color online) (a) The magnetic minor switching loop of the F$_{2}$ layer for a type B stack, where B$_{1}$ and B$_{2}$ represent the low and high switching field, respectively. (b) Resistance of the stack as function of direct current (I$_{\text{dc}}$) sent through it. The switching from the anti parallel resistance state to the parallel state and vice versa is cause by the spin-transfer torque induced by I$_{\text{dc}}$. A constant B of 40 mT is applied to ensure that we are in the middle of the minor loop, where both the parallel and anti parallel magnetization alignment constitute a stable state.}
\label{fig:minor}
\end{figure}

The magnetic minor loop measurement is presented in Fig. \ref{fig:minor}(a), where the first harmonic response is plotted as a function of B. Here we only look at the switching of the 5 nm thick F$_{2}$ layer. First M$_{1}$ and M$_{2}$ are saturated parallel by applying a high positive magnetic field. Now B is sweeped towards zero until M$_{1}$ switches, bringing the stack into the anti parallel resistance state. By reversing the B field sweep direction, before M$_{2}$ switches, a minor loop is obtained when M$_{1}$ switches back to its original parallel resistance state. The minor loop should normally be centered around B=0 but is shifted to around B=45 mT in our devices, because of the dipole field coupling between the two F layers. 

In Fig. \ref{fig:minor}(b) the STT switching experiment is given for characterization purposes. On top of the small alternating current (I$_{\text{ac}}$) of 10 $\mu$A, which gives the resistance of the stack via a lock-in detection technique, a direct current (I$_{\text{dc}}$) is sent through the stack responsible for inducing the STT. Sweeping I$_{\text{dc}}$ from -1.5 to +1.5 mA a STT switching from the parallel to anti parallel state is observed, for a positive I$_{\text{dc}}$ of 0.8 mA, and a reverse switch, for a negative I$_{\text{dc}}$ of -1.2 mA. A constant B of 40 mT is applied to make sure that we are within the minor loop (Fig. \ref{fig:minor}(a)), where both the parallel and anti parallel magnetization alignment constitute a stable state.

\begin{figure*}[t]
\includegraphics{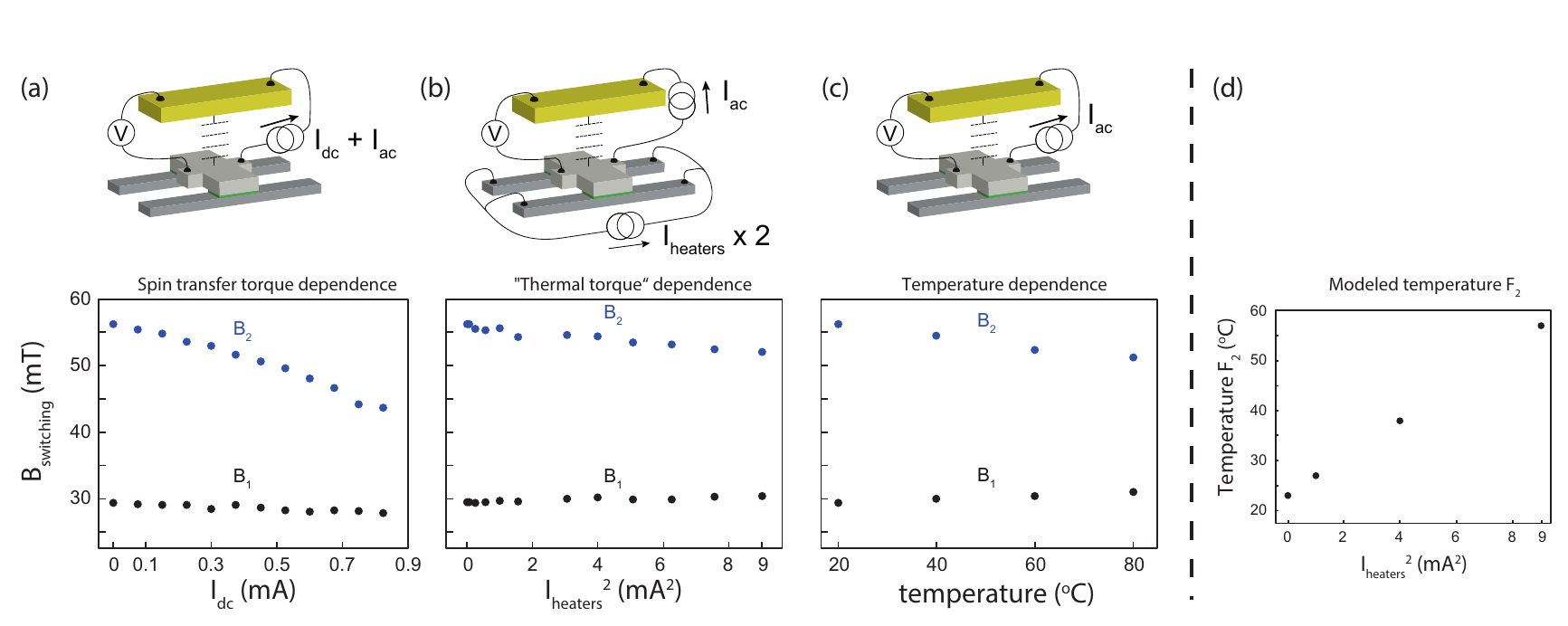}
\caption{(Color online) The evolution of the minor loop switching fields B$_{1}$ and B$_{2}$ for: (a) Spin-transfer torque, induced by sending a dc current (I$_{\text{dc}}$) through the stack. (b) ``Thermal STT'', induced by a thermal gradient across the stack by sending an I$_{\text{dc}}$ through the Pt Joule heaters. (c) Overall temperature change, induced by a controllable heater. (c) The temperature of the F$_{2}$ layer extracted from 3D finite element modeling as a function of I$_{\text{dc}}$ sent through the Pt Joule heaters.}
\label{fig:dep}
\end{figure*}

The experiments discussed above show that the switching fields B$_{1}$ and B$_{2}$ in the minor loop are changed by STT, or in other words the barrier going from the P to AP state and vice versa is changed. Measuring these two switching fields as a function of I$_{\text{dc}}$, through the F$|$N$|$F stack, therefore quantifies the response of the sample to STT, at currents below the STT switching current. Fig. \ref{fig:dep}(a) gives this evolution of B$_{1}$ and B$_{2}$, where every measurement point is an average switching field from 5 consecutively obtained minor loops. B$_{2}$ clearly shifts to lower values for higher I$_{\text{dc}}$ values, almost reaching 40 mT at an I$_{\text{dc}}$ of 0.8 mA, corresponding well to the STT switching current observed in Fig. \ref{fig:minor}(b). B$_{1}$ on the other hand only shows a very small decrease consistent with magnetic phase diagrams found for similar stacks.\cite{phase_diag}

For TSTT a similar change in B$_{1}$ and B$_{2}$ should be observed when increasing the temperature gradient across the stack. In the measurement presented in Fig. \ref{fig:dep}(b) this is investigated by determining these switching fields as a function of I$_{\text{heaters}}$, sent through the Pt Joule heaters. The results are plotted versus I$_{\text{heaters}}^{2}$ because the Joule heating scales quadratically with I$_{\text{heaters}}$. Indeed a clear quadratic decrease of B$_{2}$ is observed as one would expect for TSTT. However B$_{1}$ now seems to slightly increase, instead of showing a small decrease as seen for the STT measurement. This could indicate that the changes in B$_{1}$ and B$_{2}$ are not purely due to TSTT, but overall heating of F$_{2}$ plays an important role as well. Namely, overall heating will lower the coercive field of the F$_{2}$ layer. To further investigate this we measured the evolution of the switching fields as a function of the overall temperature of the device, without any STT or temperature gradient applied. A heating element together with a thermometer, positioned underneath and in good thermal contact with the sample, was used to controllably set the overall temperature of our device. Fig. \ref{fig:dep}(c) gives the results up to a temperature of 80$^{\text{o}}$C, showing a very similar behavior as the ``thermal'' STT dependent measurement in Fig. \ref{fig:dep}(b).

To determine if the results in Fig. \ref{fig:dep}(b) are dominated by overall heating the temperature of the F$_{2}$ layer as function of I$_{\text{heaters}}$ needs to be known. Experimentally this is difficult to determine and therefore we use our 3D finite element model, successfully used in section \ref{sec:exp_results} as well as numerous previously reported measurements.\cite{SDSE_1, SDSE_2, spin_heat_exp} The modeled temperature of F$_{2}$ versus I$_{\text{dc}}^{2}$ is given in Fig. \ref{fig:dep}(d). At an I$_{\text{heaters}}$ of 3 mA (I$_{\text{heaters}}^{2}$=9 mA$^{2}$) F$_{2}$ reaches a temperature of 57 $^{\text{o}}$C. The same I$_{\text{heaters}}$ gives a B$_{\text{switching}}$ of 52 mT, according to the measurement in Fig. \ref{fig:dep}(b), which is also found for an overall heating of $\sim$60 $^{\text{o}}$C in Fig. \ref{fig:dep}(c). In other words the change in B$_{\text{switching}}$ observed in Fig. \ref{fig:dep}(b) seems to be dominated by the overall heating of the F$_{2}$ layer.

\section{Discussion and conclusion} \label{sec:disc}

The aSDSE measurement presented in section \ref{sec:exp_results} shows that the spin heat accumulation in our devices will influence the TSTT, however this changed is assumed to be small and can effectively be neglected. Applying a temperature gradient across an F$|$N$|$F stack, presented in section \ref{sec:TT}, shows no evidence of TSTT. This we attribute to the dominance of overall heating of the magnetic layer, masking the response due to TSTT. If we indeed neglect the relatively small efficiency difference in TSTT and current driven STT, then Eq. \ref{eq:STT} and \ref{eq:TSTT} describe the torques, respectively. The $\Delta T$ needed to produce the same amount of STT, for a certain $I_{dc}$ through the stack, is then found by setting $\tau_{STT}=\tau_{TSTT}$ and gives
\begin{equation}
\Delta T =  \frac{2}{G} \frac{P_{G}}{S(P'-P_{G})} I_{dc} = \frac{P_{G}R}{S(P'-P_{G})} I_{dc},
\label{eq:Equiv_torque}
\end{equation}
where R is the resistance of the spin active part of the stack \cite{active_part} and $I_{\text{dc}}$ is the current through the stack as plotted on the x-axis in Fig. \ref{fig:dep}(a). Using R=1.3 $\Omega$ (from the 3D-FEM), S=-18 $\mu$V/K and for P and P' the values found in section \ref{sec:exp_results} we get; $\Delta T=2.9\times 10^{5}[K/A]$ $I_{\text{dc}}$. In order to switch the F$_{2}$ layer using current driven STT an $I_{dc}$ of 0.8 mA is required (see Fig. \ref{fig:minor} (b)), which then corresponds to a $\Delta T$ of 230 K, across the spin active part \cite{active_part} of the F$_{1}$ layer, for pure TSTT driven switching. It can safely be said that such a large steady state $\Delta T$ cannot be applied across such a short length and will lead to a significant increase in the background temperature. This becomes evident when determining the TSTT versus overall heating contribution in the ``thermal torque'' dependence measurement (see Fig. \ref{fig:dep}(b)). For the largest Joule heating current ($I_{heaters}$) in Fig. \ref{fig:dep}(b) B$_{\text{switching}}$ is 52 mT, which corresponds to an I$_{\text{dc}}$ of 0.375 mA for the STT dependent measurement in Fig. \ref{fig:dep}(a). The change in B$_{\text{switching}}$ observed in Fig. \ref{fig:dep}(b) would therefore need a $\Delta T$ of $110$~K, across the spin active part of the stack \cite{active_part}, if caused purely by TSTT. The model gives a $\Delta T \approx 12$ K for the largest Joule heating current, which would mean TSTT is only responsible for a maximum of $\sim$10\% of the observed B$_{\text{switching}}$ change.

In conclusion we can say that although the angle dependent measurements show that a thermal gradient will induce a TSTT, it is small and difficult to distinguish from overall heating effects. Overall heating leads to a lowering of the energy switching barrier for both the P and AP state, such that B$_{1}$ and B$_{2}$ move towards each other and gives a narrower minor loop. In the case of STT, either induced by a thermal or voltage gradient, the two switching fields should move in the same direction providing a way to distinguishing it from overall heating. Our results show that, in steady state experiments, it is difficult to avoid overall heating from being the dominant effect, unless the magnetic layer under investigation is connected directly to an almost perfect heat sink. An alternative approach would be to use use short heat pulses and look at time dependent signals as discussed in Ref. \onlinecite{aSDSE, TMR_Jia, TT_TMR} for tunnel magnetoresistance (TMR) structures. 

A combined effect of the lowering of the switching barrier by overall heating together with TSTT could of course be beneficial as the torque needed to switch will be smaller. This route is currently being investigated in the form of heat assisted switching devices.\cite{therm_ass_1, therm_ass_2, therm_ass_3} However it requires an in depth investigation and precise calibration of the timing of the two effects.

\begin{acknowledgments}
We would like to acknowledge B. Wolfs, M. de Roosz and J. G. Holstein for
technical assistance. This work is part of the research program of the
Foundation for Fundamental Research on Matter (FOM) and supported by NanoLab
NL and the Zernike Institute for Advanced Materials.
\end{acknowledgments}

\appendix

\section{aMR and aSDSE formula's} \label{app_derivation}
aMR in a symmetric F$|$N$|$F stack is described by Eq. \ref{eq:aMR}, which is found by setting $\Delta T=0$ in Eq. \ref{eq:current}. This gives:\cite{aSDSE}
\begin{align}
\frac{I_{c}(\theta)}{\Delta V}&=\frac{G}{2} \left( 1-\frac{P_{G}^{2}\tan^{2}(\theta/2)}{\eta+\tan^{2}(\theta/2)} \right) \\[12pt]
\frac{R(0)}{R(\theta)}&= \frac{\eta+(1-P_{G}^{2})\tan^{2}(\theta/2)}{\eta+\tan^{2}(\theta/2)}
\end{align}

The aSDSE is described in Eq. \ref{eq:aSdSE}, which is found by setting $I_{c}=0$ in Eq. \ref{eq:current}. This gives:\cite{aSDSE}
\begin{widetext}
\begin{align}
-\left(1 - \frac{P_{G}^{2}\tan^{2}(\theta/2)}{\eta + \tan^{2}(\theta/2)} \right) \Delta V &= \left( 1- \frac{P_{G}P'\tan^{2}(\theta/2)}{\eta + \tan^{2}(\theta/2)}\right) S\Delta T \\[12pt]
\frac{-\Delta V (\theta)}{S\Delta T} &= \dfrac{\left(\dfrac{\eta + \left(1-P_{G}P'\right)\tan^{2} \left(\theta/2\right)}{\eta + \tan^{2} \left(\theta/2\right) }\right)}{\left(\dfrac{\eta + \left(1-P_{G}^{2}\right)\tan^{2} \left(\theta/2\right)}{\eta + \tan^{2} \left(\theta/2\right) }\right)}
\end{align}
\end{widetext}

\section{Dipole magnetic field coupling} \label{app_dipole}
\begin{figure}[t]
\includegraphics{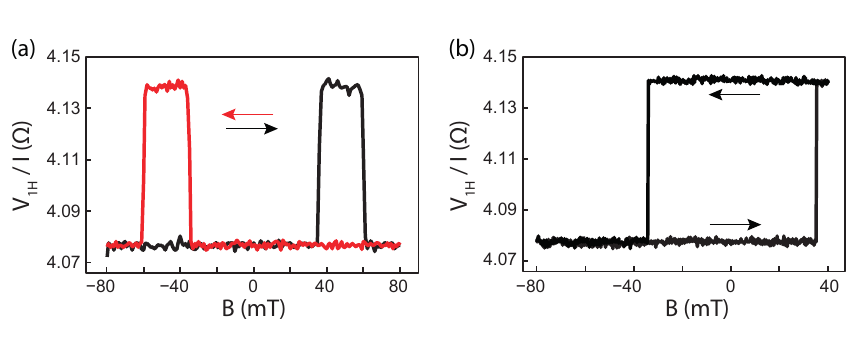}
\caption{(Color online) Measurements on device with F$|$N$|$F stack with negligible dipole field coupling (a) Resistance of the stack as a function of applied magnetic field B, spin valve measurement. (b) Minor loop switching measurement for the F$_{2}$ layer clearly showing no coupling as the loop is well centered around B=0.}
\label{fig:App}
\end{figure}

To determine the dipole field coupling between the F$_{1}$ and F$_{2}$ layer for stack type A similar devices were fabricated, with a 1.5 $\mu$m by 100 nm rectangular F$_{1}$ layer. As the F$_{1}$ layer is now much longer than the F$_{2}$ layer the dipole coupling field will become negligibly small. As the rest of the device and especially the N and F$_{2}$ layer are kept the same we are able to determine the switching field of F$_{2}$ without any coupling present. In Fig. \ref{fig:App} the spin valve and minor loop measurements are given. The minor loop is perfectly centered around B=0 confirming that the dipole field coupling is negligibly small. Furthermore we observe a switching field of 35 mT, which can be seen as the uncoupled switching field. Comparing this to the switching field of 85 mT for the coupled stacks used in the angle dependent measurements, see Fig. \ref{fig:angle_meas}, we estimate a dipole coupling field of 50 mT.

\begin{thebibliography}{20}%
\makeatletter
\providecommand \@ifxundefined [1]{%
 \@ifx{#1\undefined}
}%
\providecommand \@ifnum [1]{%
 \ifnum #1\expandafter \@firstoftwo
 \else \expandafter \@secondoftwo
 \fi
}%
\providecommand \@ifx [1]{%
 \ifx #1\expandafter \@firstoftwo
 \else \expandafter \@secondoftwo
 \fi
}%
\providecommand \natexlab [1]{#1}%
\providecommand \enquote  [1]{``#1''}%
\providecommand \bibnamefont  [1]{#1}%
\providecommand \bibfnamefont [1]{#1}%
\providecommand \citenamefont [1]{#1}%
\providecommand \href@noop [0]{\@secondoftwo}%
\providecommand \href [0]{\begingroup \@sanitize@url \@href}%
\providecommand \@href[1]{\@@startlink{#1}\@@href}%
\providecommand \@@href[1]{\endgroup#1\@@endlink}%
\providecommand \@sanitize@url [0]{\catcode `\\12\catcode `\$12\catcode
  `\&12\catcode `\#12\catcode `\^12\catcode `\_12\catcode `\%12\relax}%
\providecommand \@@startlink[1]{}%
\providecommand \@@endlink[0]{}%
\providecommand \url  [0]{\begingroup\@sanitize@url \@url }%
\providecommand \@url [1]{\endgroup\@href {#1}{\urlprefix }}%
\providecommand \urlprefix  [0]{URL }%
\providecommand \Eprint [0]{\href }%
\providecommand \doibase [0]{http://dx.doi.org/}%
\providecommand \selectlanguage [0]{\@gobble}%
\providecommand \bibinfo  [0]{\@secondoftwo}%
\providecommand \bibfield  [0]{\@secondoftwo}%
\providecommand \translation [1]{[#1]}%
\providecommand \BibitemOpen [0]{}%
\providecommand \bibitemStop [0]{}%
\providecommand \bibitemNoStop [0]{.\EOS\space}%
\providecommand \EOS [0]{\spacefactor3000\relax}%
\providecommand \BibitemShut  [1]{\csname bibitem#1\endcsname}%
\let\auto@bib@innerbib\@empty

\bibitem [{\citenamefont {Slonczewski}\ (1996)
  \citenamefont {Slonczewski} \
  and\ \citenamefont {Slonczewski}}]{Slon_STT}%
  \BibitemOpen
  \bibfield  {author} { 
  \bibinfo {author} {\bibfnamefont {J. C.}~\bibnamefont {Slonczewski}}, \ } \href {\doibase 10.1016/0304-8853(96)00062-5}
  {\bibfield  {journal} {\bibinfo  {journal} {Journ. of Magn. and Magn. Mater.}\ }\textbf {\bibinfo
  {volume} {159}},\ \bibinfo {pages} {L1-L7} (\bibinfo {year}
  {1996})}\BibitemShut {NoStop}%
\bibitem [{\citenamefont {Berger}\ (1996)
  \citenamefont {Berger} \
  and\ \citenamefont {Berger}}]{Berger_STT}%
  \BibitemOpen
  \bibfield  {author} { 
  \bibinfo {author} {\bibfnamefont {L.}~\bibnamefont {Berger}}, \ } \href {\doibase 10.1103/PhysRevB.54.9353}
  {\bibfield  {journal} {\bibinfo  {journal} {Phys. Rev. B}\ }\textbf {\bibinfo
  {volume} {54}},\ \bibinfo {pages} {9353} (\bibinfo {year}
  {1996})}\BibitemShut {NoStop}%
\bibitem [{\citenamefont {\r{A}kerman}\ (2005)
  \citenamefont {\r{A}kerman} \
  and\ \citenamefont {\r{A}kerman}}]{STT_switch}%
  \BibitemOpen
  \bibfield  {author} { 
  \bibinfo {author} {\bibfnamefont {J.}~\bibnamefont {\r{A}kerman}}, \ } \href {\doibase 10.1126/science.1110549}
  {\bibfield  {journal} {\bibinfo  {journal} {Science}\ }\textbf {\bibinfo
  {volume} {308}},\ \bibinfo {pages} {508-510} (\bibinfo {year}
  {2005})}\BibitemShut {NoStop}%
\bibitem [{\citenamefont {Valet}\ (1993)
  \citenamefont {Valet} \
  and\ \citenamefont {Valet}}]{two_channel}%
  \BibitemOpen
  \bibfield  {author} { 
  \bibinfo {author} {\bibfnamefont {T.}~\bibnamefont {Valet}},
	\bibinfo {author} {\bibfnamefont {A.}~\bibnamefont {Fert}}, \ } \href {\doibase 10.1103/PhysRevB.48.7099}
  {\bibfield  {journal} {\bibinfo  {journal} {Phys. Rev. B}\ }\textbf {\bibinfo
  {volume} {48}},\ \bibinfo {pages} {7099} (\bibinfo {year}
  {1993})}\BibitemShut {NoStop}%
\bibitem [{\citenamefont {Brataas}\ \emph {et~al.}(2000)
  \citenamefont {Brataas}, \
  and\ \citenamefont {Brataas}}]{MixC2}%
  \BibitemOpen
  \bibfield  {author} {
  \bibinfo {author} {\bibfnamefont {A.}~\bibnamefont {Brataas}}, 
  \bibinfo {author} {\bibfnamefont {Yu. V}~\bibnamefont {Nazarov}},
  \bibinfo {author} {\bibfnamefont {G. E. W.}~\bibnamefont {Bauer}}, \ } \href {\doibase 10.1103/PhysRevLett.84.2481}
  {\bibfield  {journal} {\bibinfo  {journal} {Phys. Rev. Lett.}\ }\textbf {\bibinfo
  {volume} {84}},\ \bibinfo {pages} {2481} (\bibinfo {year}
  {2000})}\BibitemShut {NoStop}%
\bibitem [{\citenamefont {Brataas}\ \emph {et~al.}(2006)
  \citenamefont {Brataas},
  \citenamefont {Bauer},
  \citenamefont {Kelly} \
  and\ \citenamefont {Brataas}}]{MixC}%
  \BibitemOpen
  \bibfield  {author} {
  \bibinfo {author} {\bibfnamefont {A.}~\bibnamefont {Brataas}}, 
  \bibinfo {author} {\bibfnamefont {G. E. W.}~\bibnamefont {Bauer}},
  \bibinfo {author} {\bibfnamefont {P. J.}~\bibnamefont {Kelly}}, \ } \href {http://www.sciencedirect.com/science/article/pii/S0370157306000238}
  {\bibfield  {journal} {\bibinfo  {journal} {Phys. Rep.}\ }\textbf {\bibinfo
  {volume} {427}},\ \bibinfo {pages} {157} (\bibinfo {year}
  {2006})}\BibitemShut {NoStop}%
\bibitem [{\citenamefont {Kovalev}}\ (2002)
  \citenamefont {Kovalev} \
  and\ \citenamefont {Kovalev}]{aMR}%
  \BibitemOpen
  \bibfield  {author} { 
  \bibinfo {author} {\bibfnamefont {A. A.}~\bibnamefont {Kovalev}},
	\bibinfo {author} {\bibfnamefont {A.}~\bibnamefont {Brataas}},
	\bibinfo {author} {\bibfnamefont {G. E. W.}~\bibnamefont {Bauer}}, \ } \href {\doibase 10.1103/PhysRevB.66.224424}
  {\bibfield  {journal} {\bibinfo  {journal} {Phys. Rev. B}\ }\textbf {\bibinfo
  {volume} {66}},\ \bibinfo {pages} {224424} (\bibinfo {year}
  {2002})}\BibitemShut {NoStop}%
\bibitem [{\citenamefont {Urazhdin}\ (2005)
  \citenamefont {Urazhdin} \
  and\ \citenamefont {Urazhdin}}]{aMR_exp}%
  \BibitemOpen
  \bibfield  {author} { 
  \bibinfo {author} {\bibfnamefont {S.}~\bibnamefont {Urazhdin}},
	\bibinfo {author} {\bibfnamefont {R.}~\bibnamefont {Loloee}},
	\bibinfo {author} {\bibfnamefont {W. P.}~\bibnamefont {Pratt, Jr.}}, \ } \href {\doibase 10.1103/PhysRevB.71.100401}
  {\bibfield  {journal} {\bibinfo  {journal} {Phys. Rev. B}\ }\textbf {\bibinfo
  {volume} {71}},\ \bibinfo {pages} {100401(R)} (\bibinfo {year}
  {2005})}\BibitemShut {NoStop}%
\bibitem [{\citenamefont {Bauer}\ (2012)
  \citenamefont {Bauer} \
  and\ \citenamefont {Bauer}}]{spin_calor}%
  \BibitemOpen
  \bibfield  {author} { 
  \bibinfo {author} {\bibfnamefont {G. E. W.}~\bibnamefont {Bauer}},
  \bibinfo {author} {\bibfnamefont {E.}~\bibnamefont {Saitoh}},
  \bibinfo {author} {\bibfnamefont {B. J.}~\bibnamefont {van Wees}},\ } \href {\doibase 10.1038/nmat3301}
  {\bibfield  {journal} {\bibinfo  {journal} {Nature Materials}\ }\textbf {\bibinfo
  {volume} {11}},\ \bibinfo {pages} {391} (\bibinfo {year}
  {2012})}\BibitemShut {NoStop}%
\bibitem [{\citenamefont {Slachter}\ (2010)
  \citenamefont {Slachter} \
  and\ \citenamefont {Slachter}}]{SDSE_1}%
  \BibitemOpen
  \bibfield  {author} { 
  \bibinfo {author} {\bibfnamefont {A.}~\bibnamefont {Slachter}},
  \bibinfo {author} {\bibfnamefont {F. L.}~\bibnamefont {Bakker}},
	\bibinfo {author} {\bibfnamefont {J-P.}~\bibnamefont {Adam}},
  \bibinfo {author} {\bibfnamefont {B. J.}~\bibnamefont {van Wees}},\ } \href {\doibase 10.1038/NPHYS1767}
  {\bibfield  {journal} {\bibinfo  {journal} {Nature Physics}\ }\textbf {\bibinfo
  {volume} {6}},\ \bibinfo {pages} {879} (\bibinfo {year}
  {2010})}\BibitemShut {NoStop}%
\bibitem [{\citenamefont {Dejene}\ (2012)
  \citenamefont {Dejene} \
  and\ \citenamefont {Dejene}}]{SDSE_2}%
  \BibitemOpen
  \bibfield  {author} { 
  \bibinfo {author} {\bibfnamefont {F. K.}~\bibnamefont {Dejene}},
	\bibinfo {author} {\bibfnamefont {J.}~\bibnamefont {Flipse}},
	\bibinfo {author} {\bibfnamefont {B. J.}~\bibnamefont {van Wees}}, \ } \href {\doibase 10.1103/PhysRevB.86.024436}
  {\bibfield  {journal} {\bibinfo  {journal} {Phys. Rev. B}\ }\textbf {\bibinfo
  {volume} {86}},\ \bibinfo {pages} {024436} (\bibinfo {year}
  {2012})}\BibitemShut {NoStop}%
\bibitem [{\citenamefont {Yu}\ (2010)
  \citenamefont {Yu} \
  and\ \citenamefont {Yu}}]{TT}%
  \BibitemOpen
  \bibfield  {author} { 
  \bibinfo {author} {\bibfnamefont {H.}~\bibnamefont {Yu}},
	\bibinfo {author} {\bibfnamefont {S.}~\bibnamefont {Granville}},
	\bibinfo {author} {\bibfnamefont {D. P.}~\bibnamefont {Yu}}, 
	\bibinfo {author} {\bibfnamefont {J.-Ph.}~\bibnamefont {Ansermet}}, \ } \href {\doibase 10.1103/PhysRevLett.104.146601}
  {\bibfield  {journal} {\bibinfo  {journal} {Phys. Rev. Lett.}\ }\textbf {\bibinfo
  {volume} {104}},\ \bibinfo {pages} {146601} (\bibinfo {year}
  {2010})}\BibitemShut {NoStop}%
\bibitem [{\citenamefont {Kovalev}\ (2006)
  \citenamefont {Kovalev} \
  and\ \citenamefont {Kovalev}}]{des_intf}%
  \BibitemOpen
  \bibfield  {author} { 
  \bibinfo {author} {\bibfnamefont {A. A.}~\bibnamefont {Kovalev}},
	\bibinfo {author} {\bibfnamefont {G. E. W.}~\bibnamefont {Bauer}},
	\bibinfo {author} {\bibfnamefont {A.}~\bibnamefont {Brataas}}, \ } \href {\doibase 10.1103/PhysRevB.73.054407}
  {\bibfield  {journal} {\bibinfo  {journal} {Phys. Rev. B}\ }\textbf {\bibinfo
  {volume} {73}},\ \bibinfo {pages} {054407} (\bibinfo {year}
  {2006})}\BibitemShut {NoStop}%
\bibitem [{\citenamefont {Hatami}\ (2007)
  \citenamefont {Hatami} \
  and\ \citenamefont {Hatami}}]{aSDSE}%
  \BibitemOpen
  \bibfield  {author} { 
  \bibinfo {author} {\bibfnamefont {M.}~\bibnamefont {Hatami}},
	\bibinfo {author} {\bibfnamefont {G. E. W.}~\bibnamefont {Bauer}},
	\bibinfo {author} {\bibfnamefont {Q.}~\bibnamefont {Zhang}},
	\bibinfo {author} {\bibfnamefont {P. J.}~\bibnamefont {Kelly}},\ } \href {\doibase 10.1103/PhysRevLett.99.066603}
  {\bibfield  {journal} {\bibinfo  {journal} {Phys. Rev. Lett.}\ }\textbf {\bibinfo
  {volume} {99}},\ \bibinfo {pages} {066603} (\bibinfo {year}
  {2007})}\BibitemShut {NoStop}%
\bibitem {active_part}%
  \BibitemOpen
  The spin active part consists of the F layers for a thickness equal to the spin relaxation length ($\lambda_{F}$) \BibitemShut {NoStop}%
\bibitem [{\citenamefont {Heikkila}\ (2010)
  \citenamefont {Heikkila} \
  and\ \citenamefont {Heikkila}}]{spin_heat}%
  \BibitemOpen
  \bibfield  {author} { 
  \bibinfo {author} {\bibfnamefont {T. T.}~\bibnamefont {Heikkil\"{a}}},
	\bibinfo {author} {\bibfnamefont {M.}~\bibnamefont {Hatami}},
	\bibinfo {author} {\bibfnamefont {G. E. W.}~\bibnamefont {Bauer}}, \ } \href {\doibase 10.1103/PhysRevB.81.100408}
  {\bibfield  {journal} {\bibinfo  {journal} {Phys. Rev. B}\ }\textbf {\bibinfo
  {volume} {81}},\ \bibinfo {pages} {100408(R)} (\bibinfo {year}
  {2010})}\BibitemShut {NoStop}%
\bibitem [{\citenamefont {Dejene}\ (2013)
  \citenamefont {Dejene} \
  and\ \citenamefont {Dejene}}]{spin_heat_exp}%
  \BibitemOpen
  \bibfield  {author} { 
  \bibinfo {author} {\bibfnamefont {F. K.}~\bibnamefont {Dejene}},
	\bibinfo {author} {\bibfnamefont {J.}~\bibnamefont {Flipse}},
	\bibinfo {author} {\bibfnamefont {G. E. W.}~\bibnamefont {Bauer}},
	\bibinfo {author} {\bibfnamefont {B. J.}~\bibnamefont {van Wees}}, \ } \href {\doibase 10.1038/NPHYS2743}
  {\bibfield  {journal} {\bibinfo  {journal} {Nature Physics}\ }\textbf {\bibinfo
  {volume} {9}},\ \bibinfo {pages} {636-639} (\bibinfo {year}
  {2013})}\BibitemShut {NoStop}%
\bibitem [{\citenamefont {Giacomoni}\ ()
  \citenamefont {Giacomoni} \
  and\ \citenamefont {Giacomoni}}]{aMR_norm}%
  \BibitemOpen
  \bibfield  {author} { 
  \bibinfo {author} {\bibfnamefont {L.}~\bibnamefont {Giacomoni}},
	\bibinfo {author} {\bibfnamefont {B.}~\bibnamefont {Dieny}},
	\bibinfo {author} {\bibfnamefont {W. P.}~\bibnamefont {Pratt, Jr.}},
	\bibinfo {author} {\bibfnamefont {R.}~\bibnamefont {Loloee}},
	\bibinfo {author} {\bibfnamefont {M.}~\bibnamefont {Tsoi}},\ } 
  \bibfield  {journal} (unpublished) \BibitemShut {NoStop}%
\bibitem [{\citenamefont {Slachter}\ (2011)
  \citenamefont {Slachter} \
  and\ \citenamefont {Slachter}}]{model}%
  \BibitemOpen
  \bibfield  {author} { 
  \bibinfo {author} {\bibfnamefont {A.}~\bibnamefont {Slachter}},
	\bibinfo {author} {\bibfnamefont {F. L.}~\bibnamefont {Bakker}},
	\bibinfo {author} {\bibfnamefont {B. J.}~\bibnamefont {van Wees}}, \ } \href {\doibase : 10.1103/PhysRevB.84.174408}
  {\bibfield  {journal} {\bibinfo  {journal} {Phys. Rev. B}\ }\textbf {\bibinfo
  {volume} {84}},\ \bibinfo {pages} {174408} (\bibinfo {year}
  {2011})}\BibitemShut {NoStop}%
\bibitem [{\citenamefont {Krivorotov}\ (2004)
  \citenamefont {Krivorotov} \
  and\ \citenamefont {Krivorotov}}]{phase_diag}%
  \BibitemOpen
  \bibfield  {author} { 
  \bibinfo {author} {\bibfnamefont {I. N.}~\bibnamefont {Krivorotov}},
	\bibinfo {author} {\bibfnamefont {N. C.}~\bibnamefont {Emley}},
	\bibinfo {author} {\bibfnamefont {A. G. F.}~\bibnamefont {Garcia}},
	\bibinfo {author} {\bibfnamefont {J. C.}~\bibnamefont {Sankey}}, 
	\bibinfo {author} {\bibfnamefont {S. I.}~\bibnamefont {Kiselev}},
	\bibinfo {author} {\bibfnamefont {D. C.}~\bibnamefont {Ralph}},
	\bibinfo {author} {\bibfnamefont {R. A.}~\bibnamefont {Buhrman}}, \ } \href {\doibase : 10.1103/PhysRevLett.93.166603}
  {\bibfield  {journal} {\bibinfo  {journal} {Phys. Rev. Lett.}\ }\textbf {\bibinfo
  {volume} {93}},\ \bibinfo {pages} {166603} (\bibinfo {year}
  {2004})}\BibitemShut {NoStop}%
\bibitem [{\citenamefont {Jia}\ (2011)
  \citenamefont {Jia} \
  and\ \citenamefont {Jia}}]{TMR_Jia}%
  \BibitemOpen
  \bibfield  {author} { 
  \bibinfo {author} {\bibfnamefont {X.}~\bibnamefont {Jia}},
	\bibinfo {author} {\bibfnamefont {K.}~\bibnamefont {Xia}},
	\bibinfo {author} {\bibfnamefont {G. E. W.}~\bibnamefont {Bauer}}, \ } \href {\doibase : 10.1103/PhysRevLett.107.176603}
  {\bibfield  {journal} {\bibinfo  {journal} {Phys. Rev. Lett.}\ }\textbf {\bibinfo
  {volume} {107}},\ \bibinfo {pages} {176603} (\bibinfo {year}
  {2011})}\BibitemShut {NoStop}%
\bibitem [{\citenamefont {Leutenantsmeyer}\ (2013)
  \citenamefont {Leutenantsmeyer} \
  and\ \citenamefont {Leutenantsmeyer}}]{TT_TMR}%
  \BibitemOpen
  \bibfield  {author} { 
  \bibinfo {author} {\bibfnamefont {J. C.}~\bibnamefont {Leutenantsmeyer}},
	\bibinfo {author} {\bibfnamefont {M.}~\bibnamefont {Walter}},
	\bibinfo {author} {\bibfnamefont {V.}~\bibnamefont {Zbarsky}},
	\bibinfo {author} {\bibfnamefont {M.}~\bibnamefont {M\"{u}nzenberg}}, 
	\bibinfo {author} {\bibfnamefont {R.}~\bibnamefont {Gareev}},
	\bibinfo {author} {\bibfnamefont {K.}~\bibnamefont {Rott}},
	\bibinfo {author} {\bibfnamefont {A.}~\bibnamefont {Thomas}},
	\bibinfo {author} {\bibfnamefont {G.}~\bibnamefont {Reiss}},
	\bibinfo {author} {\bibfnamefont {P.}~\bibnamefont {Peretzki}},
	\bibinfo {author} {\bibfnamefont {H.}~\bibnamefont {Schuhmann}},
	\bibinfo {author} {\bibfnamefont {M.}~\bibnamefont {Seibt}}, 
	\bibinfo {author} {\bibfnamefont {M.}~\bibnamefont {Czerner}},
	\bibinfo {author} {\bibfnamefont {C.}~\bibnamefont {Heiliger}}, \ } \href {\doibase : 10.1142/S2010324713500021}
  {\bibfield  {journal} {\bibinfo  {journal} {SPIN}\ }\textbf {\bibinfo
  {volume} {03}},\ \bibinfo {pages} {1350002} (\bibinfo {year}
  {2013})}\BibitemShut {NoStop}%
\bibitem [{\citenamefont {Beech}\ (2000)
  \citenamefont {Beech} \
  and\ \citenamefont {Beech}}]{therm_ass_1}%
  \BibitemOpen
  \bibfield  {author} { 
  \bibinfo {author} {\bibfnamefont {R. S.}~\bibnamefont {Beech}},
	\bibinfo {author} {\bibfnamefont {J. A.}~\bibnamefont {Anderson}},
	\bibinfo {author} {\bibfnamefont {A. V.}~\bibnamefont {Pohm}},
	\bibinfo {author} {\bibfnamefont {J. M.}~\bibnamefont {Daughton}}, \ } \href {\doibase : 10.1063/1.372720}
  {\bibfield  {journal} {\bibinfo  {journal} {J. Appl. Phys.}\ }\textbf {\bibinfo
  {volume} {87}},\ \bibinfo {pages} {6403} (\bibinfo {year}
  {2000})}\BibitemShut {NoStop}%
\bibitem [{\citenamefont {Wang}\ (2004)
  \citenamefont {Wang} \
  and\ \citenamefont {Wang}}]{therm_ass_2}%
  \BibitemOpen
  \bibfield  {author} { 
  \bibinfo {author} {\bibfnamefont {J.}~\bibnamefont {Wang}},
	\bibinfo {author} {\bibfnamefont {P. P.}~\bibnamefont {Freitas}}, \ } \href {\doibase : 10.1063/1.1646211}
  {\bibfield  {journal} {\bibinfo  {journal} {Appl. Phys. Lett.}\ }\textbf {\bibinfo
  {volume} {84}},\ \bibinfo {pages} {945} (\bibinfo {year}
  {2004})}\BibitemShut {NoStop}%
\bibitem [{\citenamefont {Xi}\ (2010)
  \citenamefont {Xi} \
  and\ \citenamefont {Xi}}]{therm_ass_3}%
  \BibitemOpen
  \bibfield  {author} { 
  \bibinfo {author} {\bibfnamefont {H.}~\bibnamefont {Xi}},
	\bibinfo {author} {\bibfnamefont {J.}~\bibnamefont {Stricklin}},
	\bibinfo {author} {\bibfnamefont {H.}~\bibnamefont {Li}},
	\bibinfo {author} {\bibfnamefont {Y.}~\bibnamefont {Chen}},
	\bibinfo {author} {\bibfnamefont {X.}~\bibnamefont {Wang}},
	\bibinfo {author} {\bibfnamefont {Y.}~\bibnamefont {Zheng}},
	\bibinfo {author} {\bibfnamefont {Z.}~\bibnamefont {Ghao}},
	\bibinfo {author} {\bibfnamefont {M. X.}~\bibnamefont {Tang}},\ } \href {\doibase : 10.1109/TMAG.2009.2033674}
  {\bibfield  {journal} {\bibinfo  {journal} {IEEE Trans. Magn.}\ }\textbf {\bibinfo
  {volume} {46}},\ \bibinfo {pages} {860} (\bibinfo {year}
  {2010})}\BibitemShut {NoStop}%
\end{thebibliography}
\end{document}